# Betweenness Centrality as a Driver of Preferential Attachment in the Evolution of Research Collaboration Networks *

Alireza Abbasi [1], Liaquat Hossain [1], and Loet Leydesdorff [2]

[1] Centre for Complex Systems Research, Faculty of Engineering and IT, University of Sydney, Sydney, NSW 2006, Australia;
[2] Amsterdam School of Communication Research (ASCoR), University of Amsterdam, Kloveniersburgwal 48, 1012 CX Amsterdam, The Netherlands;

**Abstract**

We analyze whether preferential attachment in scientific coauthorship networks is different for authors with different forms of centrality. Using a complete database for the scientific specialty of research about "steel structures," we show that betweenness centrality of an existing node is a significantly better predictor of preferential attachment by new entrants than degree or closeness centrality. During the growth of a network, preferential attachment shifts from (local) degree centrality to betweenness centrality as a global measure. An interpretation is that supervisors of PhD projects and postdocs broker between new entrants and the already existing network, and thus become focal to preferential attachment. Because of this mediation, scholarly networks can be expected to develop differently from networks which are predicated on preferential attachment to nodes with high degree centrality.

## 1 INTRODUCTION

Collaboration is one of the defining features of modern science in recent decades (Milojevic, 2010; Persson *et al*., 2004; Wagner, 2008). Although the concept is perhaps difficult to define (Woolgar, 1976), Hara *et al*. (2003) suggested that collaboration presumes at least two common elements: (1) working together for a common goal and (2) sharing knowledge. In our opinion, collaboration can be considered as a social process (Bordons & Gómez, 2000; Milojevic 2010;

---

Shrum, Genuth, & Chompalov, 2007). As Milojevic (2010 at p. 1410) formulated, "The most commonly used methods for studying collaboration networks have been: bibliometrics (Bordons & Gómez, 2000; Glänzel, 2002; Glänzel & Schubert, 2004); social network analysis/network science (Barabási *et al.*, 2002; Kretschmer, 1997; Newman, 2001c, 2004a; Wagner, 2008; Wagner & Leydesdorff, 2005); qualitative methods of observation and interviews (Hara *et al.*, 2003; Shrum *et al.*, 2007); and surveys (Birnholtz, 2006; Lee & Bozeman, 2005).

In academia, co-authorship is the most visible and accessible indicator of scientific collaboration (Abbasi, Altmann & Hwang, 2010) and "has thus been frequently used to measure collaborative activity" (Milojevic 2010), especially in bibliometric (Borgman and Furner 2002) and network-analysis studies (Milojevic 2010). Bibliometric studies of co-authorship have emphasized the effects of collaboration on scientific productivity (publications and citations) as well as on organizational and institutional aspects of collaboration applied to different units of analysis (authors, institutions, and countries) (Milojevic, 2010) (for example see Abbasi, Hossain, Uddin, Rasmussen, 2011) . On the other hand, Network studies have focused primarily on the mechanisms in the formation of "collaboration networks and understanding the underlying structures and processes leading to the observed structures" (Milojevic, 2010) (for example see Abbasi, Altmann & Hossain, 2011).

Moody (2004) indicated that authors with many collaborators and high scientific prestige gain connections from authors that are newly entering the network more than their colleagues. Recently, some studies used collaboration networks to study network dynamics (Barabási & Albert, 1999; Barabási *et al.*, 2002; Newman, 2001a) in order to reveal the existence of specific network topologies and preferential attachment as a structuring mechanism (Milojevic, 2010).

Barabási and Albert (1999) originally proposed preferential attachment as a key mechanism in the development and evolution of networks: new nodes attach preferentially to existing nodes that are already well connected; in other words, to nodes with a high degree centrality. This suggests that the evolution and expansion of networks not only depend on the growth of network (adding more nodes and links to the network) but also follows a specific ("scale-free") pattern. Whereas many networks, (e.g., World Wide Web, citation networks) follow this model of competition, it has remained a matter of debate whether the model applies to social networks (Newman, 2008; Wagner & Leydesdorff, 2005).

Scientific collaboration networks are a complex kind of social networks since both the numbers of authors (nodes) and co-authorship links among them are growing over time. Additionally, the structure of the network (the way the authors are connected) and the positions of authors in the network may vary over time. Analysis of the attachment behaviour of authors (as nodes) in terms of the nodes' positional properties may help to explore the dynamics of structural change and evolutionary behaviour in scientific collaboration networks.

In this paper, we present a study of a collaboration (co-authorship) network and investigate how authors behave during evolution (expansion) of this co-authorship network. We focus on a specific field of science—namely, research about "steel research"—with which one of us is intimately familiar, to investigate authors' attachments to specific positions in the network. The positions are indicated using centrality metrics from social network analysis studies. The three main standard centrality measures—degree, closeness and betweenness—reflect different positions and consequently roles of the actors in a network.

In other words, we hypothesize that authors attach differently to authors who are already well connected (high degree centrality), close to all others (high closeness centrality), or well bridging (brokering) between authors (high betweenness centrality). Can the general notion of "preferential attachment" thus be refined? We envisage extending this model in search of the favorable positions of researchers in their collaboration networks which give them capacity to attract more co-authors in a next stage.

During evolution of a collaboration network, attachments (new links) can happen: (1) between new authors (that is, authors added in a next period) and already existing authors; (2) among new authors; (3) among existing authors who were not connected previously; and (4) among existing authors who already had at least one previous collaboration. Our objective is to find these behavioral attachment patterns particularly identifying which characteristics of existing authors attract new authors (or cause new authors to attach to them). In particular, we investigate the following research questions:

- How do authors behave during the evolution of their collaboration network?

- Do positions (roles) of existing authors in a coauthorship network associate with the number of new authors collaborating with them at a next moment of time?
- What types of positions are most attractive for preferential attachment?

After reviewing the literature on social network analysis and preferential attachment in Section 2, we describe data sources and our collection methods in addition to the measures that will be used in Section 3. Section 4 provides the results of our analysis and finally the paper ends with conclusions and a discussion of implications of this study.

## 2 SOCIAL NETWORK ANALYSIS AND PREFERENTIAL ATTACHMENT

### 2.1 Social Network Analysis

A social network is a set of individuals or groups each of which has connections of some kind to some or all of the others. In the language of social network (that is, graph) analysis, the people or groups are called vertices, actors or nodes and the connections are edges, ties or links. Both actors and ties can be defined in different ways depending on the research questions of interest. An actor can be a single person, a team, or a company. A tie could be a friendship between two people, collaboration or common member between two teams, or a business relationship between companies.

Social network analysis has produced many results concerning social influence, social groupings, inequality, disease propagation, communication of information, and indeed almost every topic that has interested 20th century sociology (Newman, 2001b; 2004b). Social network analysis enables us to study the networks and their participants (nodes) and relations among them. Social network analysts argue that networks operate on many levels, from friends up to the level of nations. The networks play a critical role in determining the way problems are solved, organizations are run, markets evolve, and the extent to which individuals succeed in achieving their goals. Social networks have been analyzed to identify areas of strengths and weaknesses within and among research organizations, businesses, and nations as well as to direct scientific development and funding policies (Owen-Smith, Riccaboni, Pammolli, & Powell, 2002; Sonnenwald, 2007).

In a scientific collaboration network, nodes are authors and ties (links) are co-authorship relations among them. A tie exists between each two authors (scholars) if they have at least one co-authored publication. In general, scientific collaboration (co-authorship) networks can be represented as a graph. Figure 1 shows an example. The nodes (actors, vertices) of the graph represent authors and the links (ties, edges) between each two nodes indicate a co-authorship relationship between them. The weights of links denote the number of publications that two authors (co-authors) have jointly published.

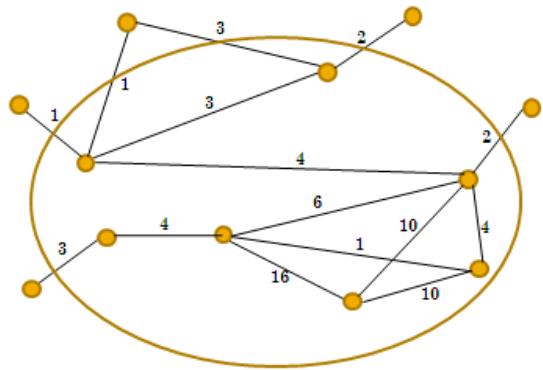

**Figure 1**. An example of co-authorship network of an academic community (adapted from Abbasi et al. (2010))

Recently, the analysis of networks and particularly the dynamics in the evolution of large networks has become of greater interest to more authors. Given the increasing evidence that networks obey unexpected scaling laws (Albert, Jeong, & Barabási, 1999; Barabási & Albert, 1999) that can be interpreted as signatures of deviation from randomness (Jeong, Néda, & Barabási, 2003), there have been efforts resulting in a class of models that view networks as evolving dynamical systems, rather than as static. These approaches, look for universalities in the dynamics governing network evolution (Jeong *et al.*, 2003).

Most models of evolving network are based on two ingredients (Barabási & Albert, 1999): growth and preferential attachment. The growth hypothesis suggests that networks tend to expand by the addition of both new nodes and links between the nodes, while the hypothesis of preferential attachment states that new nodes attach preferentially to existing (old) nodes that are already well connected. In other words, a new node is connected to some old nodes in the network based on its number of links; that is, degree centrality. These models indicate that—as Barabási & Albert (1999, at p. 509) formulated—"the development of large networks is governed by robust self-organizing phenomena that go beyond the particulars of the individual systems."

## 2.2 Preferential attachment

The "preferential attachment" process is based on the principle that "the rich get richer" or more generally "cumulative advantage." This mechanism was originally proposed by Yule (1925) (and is therefore known as leading to the Yule distribution) and is also known as "the Matthew effect" which was originally formulated by Merton (1968). The mechanism was elaborated by Price (1965 and 1976) who used the terminology of "cumulative advantage." All these processes with different names are based on a general mechanism through which a relatively favorable position can be considered as a resource to generate further gains (DiPrete & Eirich, 2006). The terminology of "preferential attachment" itself was originally used by Barabási and Albert (1999) who made the concept basic to the emerging "network science."

According to Newman (2008), the application of "preferential attachment" to processes in the new network science helps to model "a quantitative mechanism or mechanisms by which a network forms, usually in an effort to explain how the observed structure of the network arises." The specification of a dynamic mechanism makes it into one of the most important classes among the network models. The focus is thus on modeling the network generation and its evolution rather than modeling the network topology (Kronegger, Mali, Ferligoj, & Doreian, in press).

## 3 DATA AND MEASURES

### 3.1. Data Sample

For our analysis, we used a portion of a large longitudinal dataset which has been used to study the evolutionary dynamics of scientific collaboration networks of a research field indicated here as "*steel structures*" (Abbasi, Hossain, Uddin, & Rasmussen, 2011). To construct the dataset, we extracted publications using the string "*steel structure*" in the titles, keywords, or abstracts in the top 15 specified journals of the field (shortlisted by one of the authors as an expert of the field) and restricting the search to publications in English.

After extracting meta-data of these publications from Scopus [1] (one of the main sources of bibliometric data), we imported the information into a relational database. Upon comparison with our originally data, we found the affiliation information to be messy with several fields missing for some of publication and with different spellings of names of institutions, cities, and countries in the address information. Therefore, in a second step we carefully undertook manual checks (using Google) to fill out the missing fields. Additionally, we merged the universities and departments that had different names (e.g., misspellings or using abbreviations) in the originally extracted records. The database was thus made complete so that the author names are disambiguated.

For this study, we use only the publications published between 1999 and 2009. After cleaning the publication data, the resulting database contained 1,869 publications reflecting the contributions of 3,004 authors from 1,324 institutes in 77 countries.

### 3.2. Measures

A common method used to understand networks and their nodes in a static design is to evaluate the location of nodes in the network in terms of strategic positions. Node centrality concepts and measures help determine the importance of a node in a network. Bavelas (1950) was the pioneer in this field who initially investigated formal properties of centrality as a relation between structural centrality and influence in group process. He proposed several centrality concepts. Later, Freeman (1979) argued that centrality is an important structural factor influencing leadership, satisfaction, and efficiency. To quantify the importance of an actor in a social network, various centrality measures have been proposed over the years (Scott, 1991):

---

[1] at www.scopus.com.

- **Degree Centrality**

The simplest and easiest way of measuring a node's centrality is by counting the number of other nodes connected directly to this node. This "degree" of a node can be regarded as a measure of local centrality (Scott, 1991). It is worth to note that a central node is not necessarily at the center of the network physically. The degree centrality of node $k$ (i.e., $p_k$) is defined as follows:

$$C_D(p_k) = \sum_{i=1}^{n} a(p_i, p_k) \tag{1}$$

where $n$ is the number of nodes in the network and $a(p_i, p_k) = 1$ if and only if node $i$ and $k$ (i.e., $p_i$ and $p_k$) are connected; $a(p_i, p_k) = 0$ otherwise.

The concept of node centrality originated in the sociometric literature of the 'star' (Scott, 1991) which is a central node with many direct connections to other nodes. The simplest and easiest way of measuring node centrality is accordingly by the degree of the different nodes in the network. A node in a position with high degree centrality can influence the group by withholding or distorting information in transmissions (Bavelas, 1948; Freeman, 1979). Thus, degree centrality reflects the node's position and role in terms of popularity and activity of the node (Freeman, 1979) through knowing more people. Furthermore, nodes with high degree centrality could be identified as the informal leaders of the group (Krackhardt, 2010).

- **Closeness Centrality**

Freeman (1979, 1980) proposed closeness as a measure of global centrality in terms of the distances among various nodes. Sabidussi (1966) originally had suggested this concept in his work as a 'sum distance', that is, the sum of the 'geodesic' distances (the shortest path between any particular pair of nodes in a network) to all other nodes in the network. By simply calculating the sum of distances of a node to others we will have 'farness': how far the node is from other nodes. Thus, one needs to use the inverse of the farness as a measure of closeness. So, a node can be considered as globally central if it lies at the shortest distance from many other nodes; in other words, it is 'close' to many of the other nodes in the network.

In unconnected networks, every node is at an infinite distance from at least one other node, and the closeness centrality of all nodes is then 0. To solve this problem, Freeman (1979) proposed another way for calculating closeness of a node as the *"sum of the reciprocal* *distances"* of that node to all other nodes. So, closeness centrality of node *k* (i.e., $p_k$) is defined by Freeman as follows:

$$C_C(p_k) = \sum_{i=1}^{n} d(p_i, p_k)^{-1} \qquad (2)$$

where $d(p_i, p_k)$ is the geodesic distance (shortest paths) linking $p_i$ and $p_k$.

A node with the nearest position (on average) to all others can most efficiently obtain information and disseminate information quickly through the network. Thus, closeness centrality is a proxy for the independence and efficiency for communicating with other nodes in the network.

- **Betweenness Centrality**

Another global measure of centrality is betweenness which was also proposed by Freeman (1979). One considers the number of times a particular node lies 'between' the various other nodes in the network. Betweenness centrality of a node is defined as the portion of the number of shortest paths (between all pairs of nodes) that pass through the given node divided by the number of shortest path between any pair of nodes (regardless of passing through the given node) (Borgatti, 1995). More precisely, the betweenness of node *k* (i.e., $p_k$) is formulated as follows:

$$C_B(p_k) = \sum_{i<j}^{n} \frac{g_{ij}(p_k)}{g_{ij}}; i \neq j \neq k \qquad (3)$$

where $g_{ij}$ is the geodesic distance (shortest paths) linking $p_i$ and $p_j$ and $g_{ij}(p_k)$ is the geodesic distance linking $p_i$ and $p_j$ that contains $p_k$.

Nodes with high betweenness centrality play the role of a broker or gatekeeper to connect the nodes and sub-groups. So, they can most frequently control information flows in the network (Burt, 1995). Due to dependency of others on nodes with high betweenness centrality, the latter is often considered as an indicator of the power and influence these actors have in a group or organization (Krackhardt, 2010).

# 4 ANALYSIS AND RESULTS

Table 1 indicates the growth of the coauthorship network in this set by showing the number of publications, the number of authors, the number of links among them, and the average links per author during network evolution between 1999 and 2009. It also shows the cumulative number of publications, authors, and links for each year. The number of links reflects the sum of the frequency of collaborations among each pair of co-authors.

The results indicate that the growth of the number of new links is higher than the growth of the number of new authors during the period 1999-2009. The number of authors has been almost doubled (from 229 authors in 1999 to 409 authors in 2009) but the number of links has increased more than three times from 234 links in 1999 to 788 links in 2009. This increase reflects the new links (collaborations) among existing authors in each period in addition to the new links with and among the new authors. For all years after 1999, the number of links is larger than the number of authors, and the proportion of the number of new links per new author increases (albeit with fluctuations) during the period under study.

Table 1. Authors and their co-authorship links statistics over time

| New entries frequencies | | | | | Cumulative frequencies | | | | |
| --- | --- | --- | --- | --- | --- | --- | --- | --- | --- |
| Year | # of publications | # of authors | # of Links | Avg. Links/Au | Year | # of publications | # of authors | # of Links | Avg. Link/Au |
| 1999 | 117 | 229 | 234 | 1.02 | 1999-1999 | 117 | 229 | 234 | 0.96 |
| 2000 | 166 | 251 | 313 | 1.25 | 1999-2000 | 283 | 480 | 547 | 1.01 |
| 2001 | 115 | 175 | 265 | 1.51 | 1999-2001 | 398 | 655 | 814 | 1.07 |
| 2002 | 111 | 160 | 218 | 1.36 | 1999-2002 | 509 | 815 | 1030 | 1.08 |
| 2003 | 118 | 188 | 287 | 1.53 | 1999-2003 | 627 | 1003 | 1317 | 1.12 |
| 2004 | 191 | 293 | 456 | 1.56 | 1999-2004 | 818 | 1296 | 1773 | 1.17 |
| 2005 | 149 | 242 | 359 | 1.48 | 1999-2005 | 967 | 1538 | 2132 | 1.20 |
| 2006 | 220 | 366 | 690 | 1.89 | 1999-2006 | 1187 | 1904 | 2822 | 1.30 |
| 2007 | 231 | 352 | 658 | 1.87 | 1999-2007 | 1418 | 2256 | 3480 | 1.34 |
| 2008 | 226 | 339 | 649 | 1.91 | 1999-2008 | 1644 | 2595 | 4129 | 1.37 |
| 2009 | 225 | 409 | 788 | 1.93 | 1999-2009 | 1869 | 3004 | 4917 | 1.42 |

Interestingly the average number of links per author almost doubled from 1.02 (in 1999) to 1.96 (in 2009). This shows the increasing trend of collaborations among authors in this field over time. Although 2007 is the most productive year of the field (with the largest number of publications), the number of new authors, the number of new links, and also the average number of links per author is the highest in 2009. Using cumulative numbers of authors and links over time (right side of Table 1), the number of authors and links among them is increasing rapidly, and also the average number of links per author increases continuously from 0.96 in 1999 to 1.42 in 2009.

## 4.1. The Attachment Behavior of Authors and Links

In order to answer our first research question of how nodes (authors) behave during the evolution of the coauthorship network, we evaluate different forms of attachments between new and existing authors and within these two subsets. Since during evolution of the co-authorship network both new authors and new links are adding to the network, we investigate authors and links attachment behaviour first separately over time.

### 4.1.1. The attachment behaviour of authors

As an example, Figure 2 shows the co-authorship network for the year 2000 which includes existing authors in 1999 (red diamonds inside the oval)—this is the co-authorship network in the year 1999—and the newly attached authors in 2000 (blue circles). The thickness of the links is proportionate to the number of collaborations between each pair of co-authors. The co-authorship network shows just a few links between new authors and existing authors when compared to grouping among the new authors or existing authors. Thus, there are many new authors who are not connected to any of the previously existing authors.

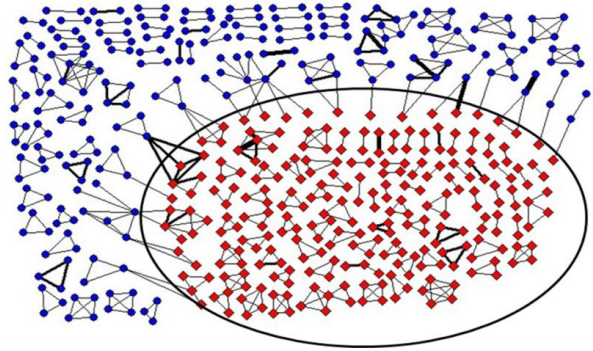

**Figure 2**. Authors' collaboration network 1999-2000 (red diamonds: existing authors in 1999; blue circles: newly added researchers in 2000)

Table 2 shows the number of authors, the number of new authors, and the number of new authors who have attached to at least: (i) another new author and (ii) an existing author. It also shows the number of existing (old) authors (up to the previous year) who have attached to at least: (i) a new author; (ii) another existing author; and (iii) any author (that is, independently of whether this is a new or old author). Since in a co-authorship network—we ignore single-authored publications—a new author will be added to the network because of connecting to either an existing author or another new author, we did not add this frequency of attachment between a new author and any author as it is by definition 100%.

Table 2. Authors (nodes) frequency over time

| Year | cumulative # of authors | # new authors | # **NEW author** attached to at least | | # **OLD authors** attached to at least | | |
|---|---|---|---|---|---|---|---|
| | | | a **NEW** author | an **OLD** author | a **NEW** author | an **OLD** author | **any** author |
| 1999 | 229 | | | | | | |
| 2000 | 480 | 251 | 238 (95%) | 27 (11%) | 19 (8%) | 24 (10%) | 36 (16%) |
| 2001 | 655 | 175 | 155 (89%) | 49 (28%) | 41 (9%) | 37 (8%) | 58 (12%) |
| 2002 | 815 | 160 | 136 (85%) | 33 (21%) | 44 (7%) | 40 (6%) | 56 (9%) |
| 2003 | 1003 | 188 | 163 (87%) | 38 (20%) | 37 (5%) | 42 (5%) | 61 (7%) |
| 2004 | 1296 | 293 | 260 (89%) | 49 (17%) | 57 (6%) | 53 (5%) | 75 (7%) |
| 2005 | 1538 | 242 | 218 (90%) | 46 (19%) | 61 (5%) | 53 (4%) | 76 (6%) |
| 2006 | 1904 | 366 | 330 (90%) | 84 (23%) | 77 (5%) | 79 (5%) | 108 (7%) |
| 2007 | 2256 | 352 | 309 (88%) | 77 (22%) | 96 (5%) | 105 (6%) | 130 (7%) |
| 2008 | 2595 | 339 | 295 (87%) | 99 (29%) | 99 (4%) | 131 (6%) | 145 (6%) |
| 2009 | 3004 | 409 | 373 (91%) | 111 (27%) | 92 (4%) | 111 (4%) | 131 (5%) |

For example, the last row of Table 2 shows that 373 of the newly attached authors in 2009 (out of 409 authors) have been attached to at least another new author. But only 92 existing authors (out of 2595 authors up to 2008) have been attached to at least one of the 409 newly attached authors in 2009.

The results show that a minority of the new authors attach to existing authors, while most of them attach to other new authors. Studying attachment behaviour of existing authors, interestingly shows relatively few of them connect to any other authors (no matter whether s/he is a new or existing author). This rate decreases as the network grows over time. The attachments of existing authors, however, are almost equally attached to other existing authors or new authors.

### *4.1.2. Different types of new links (attachments)*

During the evolution of a co-authorship network, several types of new links may form: (i) among new authors; (ii) between new authors and existing authors; (iii) among existing authors who had no collaboration (link) before; and (iv) among existing authors already linked (i.e., the new links at time t among existing authors who had at least one collaboration at any time before t). Table 3 shows the number of links (collaborations) and new links per year followed by the frequency of four different types of new links.

Table 3. Different links' types frequency over time

| Year | cumulative # of links | # of new links | # of new links | | | |
|---|---|---|---|---|---|---|
| | | | among NEW authors | between NEW & OLD authors | among OLD authors not connected | among OLD co-authors |
| 1999 | 234 | | | | | |
| 2000 | 547 | 313 | 260 (83%) | 35 (11%) | 0 (0%) | 18 (6%) |
| 2001 | 814 | 267 | 76 (35%) | 87 (33%) | 1 (0%) | 40 (15%) |
| 2002 | 1030 | 216 | 72 (39%) | 73 (34%) | 1 (0%) | 30 (14%) |
| 2003 | 1317 | 287 | 65 (27%) | 69 (24%) | 9 (3%) | 33 (11%) |
| 2004 | 1773 | 456 | 104 (26%) | 123 (27%) | 12 (3%) | 32 (7%) |
| 2005 | 2132 | 359 | 103 (32%) | 104 (29%) | 7 (2%) | 32 (9%) |
| 2006 | 2822 | 690 | 150 (24%) | 165 (24%) | 21 (3%) | 39 (6%) |
| 2007 | 3480 | 658 | 170 (31%) | 190 (29%) | 18 (3%) | 75 (11%) |
| 2008 | 4129 | 649 | 202 (37%) | 218 (34%) | 27 (4%) | 72 (11%) |

| 2009 | 4917 | 788 | 207 (30%) | 225 (29%) | 32 (4%) | 62 (8%) |

The results indicate that most of the new links (attachments) occur among newly added authors and then between new and existing authors, and among existing (old) co-authors, respectively. Although very few disconnected existing authors attach to each other, existing co-authors tend to coauthor more frequently. It follows from Table 3 that existing authors collaborate either to new authors (most possibly in supervision relations) or with previous coauthors. The previous collaboration can be expected to have generated trust among these authors which in turn facilitates their new collaboration.

The number of new links among new authors and between new and existing authors is almost equal with the exception of 2000 when there was an exceptionally high number of links among new authors.

### 4.2. Preferential Attachment Behavior during Network Evolution

In order to answer our second and third research questions about the behaviours of new authors' attachments to existing authors based on their positional characteristics, first we calculate existing centrality measures (i.e., degree, closeness and betweenness for each year) for all authors, and correlate these values with the frequencies of new authors and links attached to them in the year thereafter. Using Spearman rank correlations, we measure the correlations between existing authors' centrality measures and the numbers of attached authors and links to them in the following year between 1999 and 2009.

Table 4 shows that the existing authors' centrality measures positively and significantly correlate with the numbers of new authors attaching to them (except for degree and closeness centrality measures in 1999 and 2000). Results of the correlation test not only support the preferential attachment process for this scientific collaboration network—new authors prefer to attach to well-connected authors (having high degree centrality)—but also asserts that the new authors prefer to attach to the authors who are close to all other authors in the co-authorship network (having high closeness centrality) and the authors who entertain the role of brokering (and bridging) in the network (having high betweenness centrality).

Table 4. Spearman correlation between the existing authors' centrality measures and their attachment frequency in each period

| Centrality Measures | | number of **new authors** attached to the existing authors in the next year | | | | | | | | | |
|---|---|---|---|---|---|---|---|---|---|---|---|
| | | **1999** | **2000** | **2001** | **2002** | **2003** | **2004** | **2005** | **2006** | **2007** | **2008** |
| **Number of authors** | | 229 | 480 | 655 | 815 | 1003 | 1296 | 1538 | 1904 | 2256 | 2595 |
| **Degree Centrality** | P | .104 | .082 | **.149**\*\* | **.139**\*\* | **.126**\*\* | **.191**\*\* | **.164**\*\* | **.146**\*\* | **.148**\*\* | **.155**\*\* |
| **Closeness Centrality** | P | .091 | .082 | **.175**\*\* | **.148**\*\* | **.089**\*\* | **.132**\*\* | **.123**\*\* | **.105**\*\* | **.111**\*\* | **.118**\*\* |
| **Betweenness Centrality** | P | **.170**\*\* | **.104**\* | **.370**\*\* | **.232**\*\* | **.233**\*\* | **.322**\*\* | **.296**\*\* | **.292**\*\* | **.246**\*\* | **.273**\*\* |

\* Correlation is significant at the 0.05 level (2-tailed).
\*\* Correlation is significant at the 0.01 level (2-tailed).

We also examined the correlations between the existing authors' centrality measures and the number of *links* attached to them. The number of new links attach to an existing author considers both the number of newly attached authors and also recurrent collaborations with other existing nodes (no matter the other existing author was connected before or not). However, the results of correlation test were virtually similar to those in Table 4.

The results reveal that the correlation between betweenness centrality of existing authors and their attachment frequency (the number of new authors and links attached to them) in the following year is always significant and much higher than the degree and closeness centrality measures during the evolution of this collaboration network over time. In other words, authors with high betweenness centrality attract more new co-authors than the well-connected authors or the authors who are close to all others.

It is worth to note that looking at each centrality measure values over time, the correlation between the number of newly attached authors and degree centrality remains almost constant (with some fluctuation). But for closeness centrality, the correlation is fluctuating and for betweenness centrality it is increasing over time. Therefore, we may infer that as the collaboration network grows, betweenness centrality becomes increasingly important for attachments or, in other words, authors with high betweenness centrality gain more power and influence to attract new co-authors. An increasing number of authors prefer to attach to the existing authors who are controlling the flow of information (communication) by having a brokering (or bridging) role in the collaboration network.

In Table 5, we provide the (rank-order) correlations between existing authors' centrality measures for each year and their number of co-authors in the next year. The results indicate that the correlation coefficient is highest for degree centrality (as expected because of the accumulative design) followed by closeness centrality and thereafter betweenness centrality. These results show that authors who have a larger number of co-authors can be expected to have many co-authors in the following period. Furthermore, authors with higher closeness centrality measures can be expected to have more co-authors (in the following period) than the authors with higher betweenness centrality measure.

Table 5. Spearman correlation between the existing authors' centrality measures and their co-author frequency in each period

| Centrality Measures | | number of (all) **co-authors** of the existing authors in the next year | | | | | | | | | |
|---|---|---|---|---|---|---|---|---|---|---|---|
| | | 1999 | 2000 | 2001 | 2002 | 2003 | 2004 | 2005 | 2006 | 2007 | 2008 |
| Number of authors | | 229 | 480 | 655 | 815 | 1003 | 1296 | 1538 | 1904 | 2256 | 2595 |
| Degree Centrality | P | .913** | .907** | .927** | .946** | .954** | .970** | .974** | .978** | .980** | .982** |
| Closeness Centrality | P | .808** | .750** | .740** | .725** | .698** | .683** | .677** | .659** | .630** | .603** |
| Betweenness Centrality | P | .170** | .446** | .492** | .508** | .486** | .493** | .498** | .488** | .482** | .487** |

\* Correlation is significant at the 0.05 level (2-tailed).
\*\* Correlation is significant at the 0.01 level (2-tailed).

While the coefficient values follow an increasing trend over time for degree centrality, this trend is negative for closeness and is approximately stable for betweenness centrality (ignoring the first period). One can consider that authors' degree centrality is always increasing over the evolution of their co-authorship network considering an accumulative design. But authors' closeness and betweenness measures vary in each period as these are global measures which depend on the topology of the network in each period. Therefore, it follows from the design that the correlation coefficients for degree centrality are very high and increasing.

Furthermore, in order to see how the position of existing authors in terms of these centrality measures have an impact on the number of newly attached authors to them, we compared the average numbers of new authors relative to authors with low and high centrality measures. The mean of each centrality measure is used as a threshold for dividing authors into two categories having low or high centrality measure. Table 6 shows the average number of new authors for each category in each period.

Table 6. Comparing authors' average number of new authors attached to the existing authors in each period

| Centrality Measures | | Average (mean) of the number **new authors** attached to the existing authors in the next year | | | | | | | | | |
|---|---|---|---|---|---|---|---|---|---|---|---|
| | | 1999 | 2000 | 2001 | 2002 | 2003 | 2004 | 2005 | 2006 | 2007 | 2008 |
| **Degree Centrality** | Low | .08 | .14 | .06 | .04 | .08 | .03 | .05 | .06 | .04 | .03 |
| | High | .17 | .23 | .24 | .18 | .17 | .18 | .20 | .15 | .16 | .15 |
| **Closeness Centrality** | Low | .11 | .15 | .05 | .06 | .09 | .06 | .05 | .07 | .05 | .05 |
| | High | .17 | .18 | .20 | .12 | .13 | .13 | .20 | .15 | .20 | .18 |
| **Betweenness Centrality** | Low | .11 | .15 | .06 | .06 | .09 | .07 | .08 | .07 | .06 | .05 |
| | High | .44 | .29 | .57 | .39 | .37 | .33 | .57 | .58 | .79 | .85 |

The results in Table 6 show that authors with high values for degree centrality have on average larger number of new co-authors compared to low-degree authors. This is the same for the authors with high closeness and betweenness. Authors with on average high betweenness centrality values have the largest number of new co-authors in each period. This effect increases over time.

Furthermore, authors with low degree centrality have the lowest number of new co-authors (on average) in each period. These results not only confirm our previous findings that the betweenness centrality of authors is more important than their degree or closeness, but also show a large gap in preferential attachment between authors with high betweenness centrality, on the one side, and high degree or closeness centrality, on the other.

# 5   DISCUSSION AND CONCLUSION

In order to investigate the attachment behaviour of authors during the temporal evolution of their co-authorship networks, we examined whether central positions (and roles) in the collaboration network generate further gains in attractivity to new nodes. Network science has introduced centrality measures as proxies for specific positions and roles of the nodes in a network. In this study of the evolution of the co-authorship relations among researchers in "*steel structure*" (between 1999 and 2009) we assessed the extent to which the main centrality measures (i.e., degree, closeness and betweenness) associated with the expectation of new co-authorships.

The results show that all three centrality values of existing authors correlate to the attachment frequency of new authors to them. However, more authors prefer to attach to authors who have higher betweenness centrality rather than those with higher degree or closeness centrality. In other words, during network evolution existing authors who have the power of controlling the communication and information flow (that is, higher betweenness centrality), attract more new co-authors than authors who have more co-authorship links (degree centrality) or those who have more direct connection to all other nodes in the network (closeness centrality).

Our results also indicate that a relatively small number of new authors attach to the existing authors. During the evolution of scientific collaboration networks, new authors not necessarily attach to existing authors and thus add to the coherence in the network. We find that also few existing authors have a new collaboration (to any author) in the following year (on average 8%); this percentage is almost equal to that for new authors. Existing authors prefer to coauthor with others with whom they already coauthored before. Furthermore, we found that authors rarely initiate coauthorship relations with other authors who have already published in their domain, but prefer to have collaboration with new authors. Probably, a large proportion of this category consists of collaborations between supervisors and their students. In a co-authorship network, authors with high betweenness centrality seem to be supervisors, since they usually have publications with both other colleagues and their graduate students during their academic life. This gives them the brokering (bridging) role to connect students to other colleagues or students.

One of our questions was to examine whether preferential attachment played a role in the temporal evolution of scientific collaboration networks following previous studies in which preferential attachment was found in the evolution of a large number of network types, including scientific collaboration networks. Whereas this was shown, our main novel contribution, however, is that we found another property of authors than their degree, namely, the brokering role, based on their position in their of co-authorship networks (i.e., betweenness centrality) to be driving the coauthorship network. Betweenness centrality indicates the preferential attachment process among authors during the network evolution more than the number of links one has accumulated (i.e., degree centrality).

A limitation of this study remains that we studied the single case of one field, namely research about "*steel structures*". Our contribution, therefore, provides mainly a hypothesis. In order to generalize these findings, one would need to investigate other scientific domains. If this relationship between betweenness centrality in coauthorship networks and new entrants to the field is systematic, it might help policy and decision makers to identify key actors who facilitate the flow of information by attaching new actors during the evolution and expansion of the networks. This helps to control the distribution of resources, information dissemination, and propagation based on a typology of network positions.

**ACKNOWLEDGMENT**

The authors appreciate the anonymous reviewers for their positive and useful comments on the early drafts of this paper.